\documentclass[conference]{IEEEtran}
\IEEEoverridecommandlockouts
\usepackage{cite}
\usepackage{amsmath,amssymb,amsfonts}
\usepackage{algorithmic}
\usepackage{graphicx}
\usepackage{textcomp}
\usepackage{xcolor}
\usepackage{comment}

\usepackage{bm}%%添加的
\usepackage{subfigure}
\usepackage{colortbl}
\usepackage{algorithm}
\usepackage{amsmath,cases}

\def\BibTeX{{\rm B\kern-.05em{\sc i\kern-.025em b}\kern-.08em
    T\kern-.1667em\lower.7ex\hbox{E}\kern-.125emX}}
\begin{document}
% 全局设置lineskip和displayskip
\setlength{\lineskiplimit}{0pt}
\setlength{\lineskip}{0pt}
\setlength{\abovedisplayskip}{6pt}   % 公式上下间隔设置为6pt比较好看，可以自己更改

\setlength{\abovedisplayshortskip}{6pt}
\setlength{\belowdisplayshortskip}{6pt}

\title{Orbital Angular Momentum Active Anti-Jamming in Radio Wireless Communications
\thanks{This work is supported in part by the National Natural Science Foundation of China under Grant 62201427 and the Natural Science Basic Research Program of Shaanxi under Grant 2024JC-YBQN-0642. \textit{(Corresponding author: Liping Liang.)}}}

\author{\IEEEauthorblockN{Kexin Zheng$^{\dagger}$, Wenchi Cheng$^{\dagger}$, and Liping Liang$^{\dagger}$}\\[0.2cm]
\vspace{-10pt}

\IEEEauthorblockA{$^{\dagger}$State Key Laboratory of Integrated Services Networks, Xidian University, Xi'an, China\\
E-mail: \{\emph{kexinzheng@stu.xidian.edu.cn}, \emph{wccheng@xidian.edu.cn}, \emph{liangliping@xidian.edu.cn}\}}

\vspace{-20pt}
}

\maketitle

\begin{abstract}
Orbital angular momentum (OAM), providing the orthogonality among different OAM modes, has attracted much attention to significantly increase spectrum efficiencies (SEs) and enhance the anti-jamming results of wireless communications. However, the SE of wireless communications is severely degraded under co-frequency and co-mode hostile jamming. Focused on this issue, we propose a novel OAM active anti-jamming scheme to significantly enhance the anti-jamming results of wireless communications under broadband hostile jamming. Specifically, the OAM transmitter with energy detection senses jamming signals to identify which OAM modes are jammed and unjammed. Based on the recognition of OAM modes, useful signals are modulated by reflecting the received co-frequency and co-mode jamming signals with the assistance of a programmable gain amplifier (PGA) to the OAM receiver, thus utilizing both the OAM modes jammed by hostile attacks and the energy of jamming signals. Meanwhile, the unjammed OAM modes allocated with total transmit power are multiplexed for useful signal transmission. Numerical results demonstrate that our proposed OAM active anti-jamming scheme can achieve high OAM mode utilization and significantly increase the SEs.

\end{abstract}

\begin{IEEEkeywords}
Orbital angular momentum (OAM), active anti-jamming, spectrum efficiency.
\end{IEEEkeywords}

\section{Introduction}
\IEEEPARstart{D}{ue} to the open and broadcast properties of wireless channels, wireless communication systems are highly vulnerable to jamming and eavesdropping attacks, thus resulting in poor quality of wireless communication systems. This problem has promoted the research on physical-layer security to protect communication messages from being jammed or wiretapped \cite{1}, \cite{1a}. Current works related to physical-layer security mainly focus on frequency hopping (FH), where the carrier frequency is quickly changed according to a pre-set hopping pattern between legitimate transceivers to suppress jamming signals\cite{2}. However, FH becomes ineffective against partial band/broadband jamming within limited bandwidth resource. Therefore, it is highly required to explore new technologies to achieve robust anti-jamming results of wireless communications under broadband jamming attacks.

Fortunately, orbital angular momentum (OAM), which describes the helical phase front of electromagnetic waves, offers a significant means of improving physical-layer security and increasing the spectrum efficiencies (SEs) of wireless communications without requiring additional resources, such as time, frequency, power, and code. One of most common methods of generating multiple OAM beams is through a uniform-circular-array (UCA) based antenna\cite{3},\cite{q}. The number of OAM modes theoretically is infinite, whereas in practice it is limited by the number of arrays. The authors of \cite{3b} and \cite{a5} pointed out that OAM multiplexing can achieve high SEs of wireless communications due to the orthogonality among different integer OAM modes. The authors of \cite{a} pointed out that the anti-jamming performance of wireless communications can be improved by utilizing OAM-based hybrid duplex cognitive frequency hopping, where the primary and secondary users in a cognitive radio network respectively employ half-duplex co-awareness and full-duplex transmission modes. To better exploit the orthogonality of OAM modes, the authors of \cite{a7} have proposed mode hopping schemes for anti-jamming, which change the OAM modes within one period fast. However, legitimate receivers cannot successfully distinguish wanted signals and jamming signals with existing anti-jamming techniques under co-frequency and co-mode jamming. Consequently, it is imperative to develop effective anti-jamming techniques to enhance the performance of wireless communications.

Ambient backscatter communication, where the backscatter transmitters modulate signals by adjusting their antenna impedance to either absorb or reflect ambient radio frequency signals and then send the modulated signals to a backscatter receiver, is regarded as a promising technology for anti-jamming\cite{a8}. The authors of\cite{10} also verified that the ambient backscatter technique can enhance the anti-jamming performance by utilizing the deep reinforcement learning algorithm.

Inspired by the concept of ambient backscatter communication, in this paper, we propose an active anti-jamming scheme which does not resist jamming but uses jamming signals as carriers for data transmission. Specifically, the OAM transmitter performs energy detection to identify the jammed and unjammed OAM modes. Based on the identification results, the OAM transmitter employs a programmable gain amplifier (PGA) to re-modulate co-mode jamming signals as carriers to send useful information. Assigned with total transmit power, the unjammed OAM modes are used to transmit multiple signals in parallel to increase the SE of wireless communication. Numerical results show that our proposed scheme can significantly enhance the SEs compared with the jammed UCA system without jammed OAM modes utilization.

The remainder of this paper is organized as follows. Section II gives the system model of our proposed OAM active anti-jamming scheme. Section III presents the OAM active anti-jamming scheme. Section IV numerically evaluates the performance of our proposed scheme. The paper concludes with Section V.

\section{System Model}\label{sec:sys}
\begin{figure}[htbp]
%\vspace{-0.4cm}%%调整标题与文字距离间距
\centerline{\includegraphics[scale=0.73]{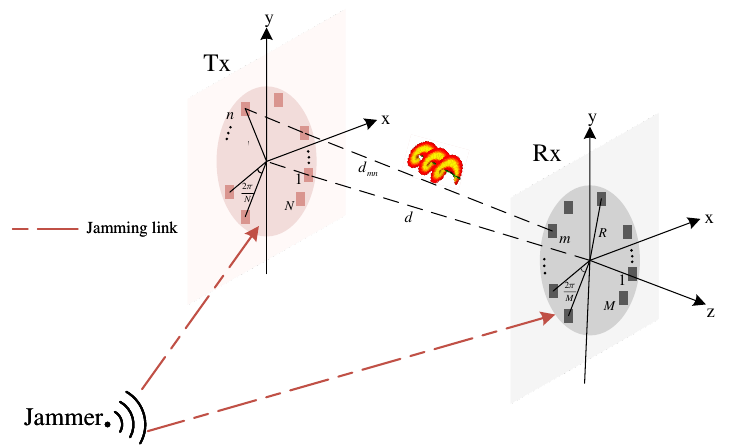}}
%\centerline{\includegraphics[scale=0.7]{UCA.eps}}
\caption{The system model of OAM active anti-jamming.}
\label{system model}
%\vspace{-0.3cm}%%调整标题与文字距离间距
\end{figure}

%\subsection{System Model}
%%%%%%%%%%%%%%%%%%%%%%%%%%%%%%%%%%%%%%%%%%%%%%%%%%%%%%%%%%%%%%%%%%
In this section, we build the OAM active anti-jamming system model as shown in Fig. \ref{system model}, where a jammer attempting to interrupt the legitimate communication between the UCA-based transmitter (Tx) and receiver (Rx) sends co-frequency and co-mode jamming signals with a high coverage range. Each element of transmit UCA is fed with the same input signals, but there is a phase delay of $2\pi l /N$ between two adjacent elements, where $l\left( \lfloor \frac{2-N}{2} \rfloor \leqslant l\leqslant \lfloor \frac{N}{2} \rfloor\right) $ denotes the order of OAM mode and $\lfloor \cdot \rfloor$ denotes the floor function. We denote $d$ the distance from the center of the transmit UCA to the center of the receive UCA, $d_{mn}$ the distance between the $n$-th $\left(1\leqslant n\leqslant N \right) $ transmit element and the $m$-th $\left(1\leqslant m\leqslant M \right) $ receive element, $r$ the radius of the transmit UCA, and $R$ the radius of the receive UCA.

As shown in Fig. \ref{system model}, we mainly consider the common high-power jamming signals carried by plane electromagnetic waves which can be considered as verticose electromagnetic waves with zero OAM modes. To enhance the anti-jamming results of wireless communications, the Tx equipped with a PGA operates in full-duplex mode. By mapping useful messages into different amplification factors, the PGA modulates the received jamming signals forwarded to the Rx. By fully leveraging the orthogonality of OAM modes, the transmitter also transmits useful signals with its power to the legitimate receiver.

\section{Proposed Active Anti-jamming Scheme}\label{sec:ASR}
%\subsection{Problem Formulation}
In this section, we propose the OAM active anti-jamming scheme. We first use energy detection to identify the jammed and unjammed OAM modes. For the jammed OAM modes, the OAM-based PGA scheme is developed to employ the jamming signals as carriers to transmit useful information. Then, the multiplexing transmission scheme between transmit UCA and receive UCA is applied to the remaining OAM modes.

\subsection{OAM Mode Identification}\label{AA}
The received jamming signal, denoted by $x_{\mathrm{J},n}\left(t \right) $, at the $n$-th element of the transmitter follows $x_{\mathrm{J},n}\left(t \right)\sim \mathcal{C} \mathcal{N} \left( 0,\sigma^2_{\mathrm{j},n} \right)$, where $t$ is the time variable. Then, with a discrete Fourier transform (DFT) algorithm at the transmitter, the decomposed OAM signals corresponding to the OAM mode $l$, denoted by $\mathrm{T}_{l}\left(t \right) $, is given by
\begin{equation}
\mathrm{T}_{l}\left(t \right) =\frac{1}{\sqrt{N}}\sum_{n=1}^N{x_{\mathrm{J},n}\left(t \right) e^{-j\frac{2\pi \left( n-1 \right)}{N}l}}.
\label{1}
\end{equation}

Then, the energy detection is performed on the decomposed signals $\mathrm{T}_{l}\left(t \right)$ by the transmitter to identify the jammed and unjammed OAM modes. We first take a $K$-point sampling on the received signal $\mathrm{T}_{l}\left(t \right)$, with a sampling interval of $T_s$. Based on Eq. (\ref{1}), the sampled signal at the $k$-th $\left(k=1,2,\cdots, K \right)$ sampling point is given by
\begin{equation}
\mathrm{T}_{l}[k]=\frac{1}{\sqrt{N}}\sum_{n=1}^N{x_{\mathrm{J},n}\left( k \right) e^{-j\frac{2\pi \left( n-1 \right)}{N}l}},
\end{equation} 
where $x_{\mathrm{J},n}\left( k \right)$ is the $k$-th point sampling value of the received jamming signal $x_{\mathrm{J},n}\left( t \right)$ and $\mathrm{T}_l[k]\sim \mathcal{C}\mathcal{N}\left(0,\sigma^2_l \right) $ follows complex Gaussian distribution with variance $\sigma^2_l=\sigma^2_{\mathrm{j},n}$. Then, the energy of the signal on the OAM mode $l$, denoted by $E_l$, is given by
\begin{equation}
E_l=\frac{1}{K}\sum_{k=1}^{K}\left|\mathrm{T}_{l}\left[k\right]\right|^2,
\label{El}
\end{equation}
where $E_l\sim \text{Ga}\left(K,\frac{\sigma^2_l}{K} \right) $ follows Gamma distribution. 

After the energy detection, we define an energy threshold ${E}_{\text{th}}$ to determine which OAM modes are jammed or unjammed. 

When $E_l\ge {E}_{\text{th}}$, the OAM mode is judged to be jammed, thus forming the set $\mathcal{L}_{\mathrm{j}}=\left\{ l\left| E_l\geqslant {E}_{\text{th}} \right. \right\} $. Otherwise, we have the unjammed OAM-mode set $\mathcal{L}_{\mathrm{u}}=\left\{ l\left| E_l < {E}_{\text{th}} \right. \right\} $.

To evaluate the accuracy of energy detection, we denote the probability of correct detection of jammed and unjammed OAM modes by $P_{\text{j}}$ and $P_{\text{u}}$, respectively. Based on Eq. (\ref{El}), we have
\begin{numcases}{}
	P_{\text{j}}=1 - F\left(E_{\text{th}};K,\frac{\sigma^2_l}{K}\right); \nonumber \\
	P_{\text{u}}=F\left(E_{\text{th}};K,\frac{\sigma^2_l}{K}\right), \label{Pd}
\end{numcases}
where $F$ is the cumulative distribution function (CDF) of $E_l$ and it is calculated by
\begin{equation}
F\left(E_{\text{th}};K,\frac{\sigma^2_l}{K}\right)=\frac{1}{\Gamma\left(K\right)}\int_{0}^{E_{\text{th}}}t^{K-1}e^{-\frac{tK}{\sigma^2_l}}dt.
\label{F}
\end{equation}

In Eq. (\ref{F}), the gamma function, denoted by $\Gamma\left(K \right) $, is given as follows:
\begin{equation}
\Gamma\left(K \right)=\int_{0}^{+\infty}x^{K-1}e^{-K}dK. 
\end{equation}

\subsection{Signal Transmission and OAM Mode Decomposition}
The initial signals for the OAM mode $l$ at the transmitter is denoted by $s_l\left(t \right)$. Then, the emitted modulated signal for the $n$-th transmit element for all OAM modes, denoted by $x_n\left(t \right)$, can be expressed as follows:
\begin{equation}
x_n\left(t \right) =\frac{1}{\sqrt{N}}\sum_{l=\lfloor \frac{2-N}{2} \rfloor}^{ \lfloor \frac{N}{2} \rfloor}s_l\left(t \right)e^{j\frac{2\pi\left(n-1\right)}{N}l}.
\label{xn}
\end{equation}

Due to the existence of jamming signals, the transmitter is required to modulate the jamming signals as carriers to emit useful signals with the jammed OAM modes, thus improving the anti-jamming performance of wireless communications. Thereby, we have 
\begin{numcases}{s_l\left( t\right) =}
		A\left( t\right)\mathrm{T}_l\left( t\right), l\in L_{\mathrm{j}}; \nonumber \\
	\hat{s}_l\left( t\right), l\in \mathcal{L}_{\mathrm{u}}, \label{sl}
\end{numcases}
where $A\left( t\right)\in\left\{a_b\right\}, b=0, 1, \cdots, B-1$ with the distribution $p\left(A\left(t \right)=a_b  \right)=p_b, \sum\limits_{b=0}^{B-1}p_b=1 $ denotes the amplification factor with the PGA corresponding to the messages to be sent by the transmitter and $\hat{s}_l\left(t \right) $ is the initial signals for the unjammed OAM modes. We assume that the transmit signals are binary sequences "0" and "1", where $B=2$ and the amplification factors with respect to "0" and "1" are denoted by $a_0$ and $a_1$, respectively.  

This paper mainly considers the line-of-sight (LoS) transmission in free space. The channel amplitude gain, denoted by $h_{mn}$, from the $n$-th element of the transmit UCA to the $m$-th element of the receive UCA, is given by
\begin{equation}
h_{mn}=\frac{\beta \lambda e^{-j\frac{2\pi}{\lambda}d_{mn}}}{4\pi d_{mn}},
\label{hmn}
\end{equation}
where $\lambda$ is the wavelength and $\beta$ represents all relevant constants, such as the divergence of vorticose electromagnetic waves and phase rotation due to the OAM mode effects of the transmit and receive antennas.

With the geometric mathematical method, the distance $d_{mn}$ can be derived as follows:
\begin{align}\label{dmn}
        d_{mn}=\sqrt{d^2+r^2+R^2-2rR\cos\left(\phi _n-\psi _m \right) }\nonumber\\
        \approx \sqrt{d^2+r^2+R^2}-\frac{rR\cos\left(\phi_n-\psi_m \right)}{\sqrt{d^2+r^2+R^2}},
\end{align}
where $\phi_n=2\pi\left(n-1 \right) /N$ and $\psi_m=2\pi\left(m-1 \right) /M$ represent the azimuthal angles of the $n$-th and $m$-th element in the transmit and receive UCA, respectively. Associating Eqs. (\ref{hmn}) and (\ref{dmn}), the channel gain $h_{mn}$ can be derived as follows:
\begin{equation}
	h_{mn} =\displaystyle\frac{\beta \lambda e^{-j\frac{2\pi}{\lambda}\sqrt{d^2+r^2+R^2}}}{4\pi d}e^{j \frac{2\pi rR}{\lambda \sqrt{d^2+r^2+R^2}}\cos \left( \phi _n-\psi _m \right) }. %第2行公式编号
\end{equation}

Then, the received signal at the $m$-th element, denoted by $y_m\left(t \right)$, can be derived as follows:
\begin{align}\label{ym}
        y_{m}\left(t \right) &=\frac{1}{\sqrt{M}}\sum_{n=1}^{N}h_{mn}x_n\left(t \right)+z_m\left(t \right)+x_{\mathrm{J},m}\left(t \right) \nonumber\\
                            &= \frac{1}{\sqrt{MN}}\sum_{l\in \left\{\mathcal{L}_{\mathrm{j}},\mathcal{L}_{\mathrm{u}}\right\}}{\sum_{n=1}^N{h_{mn}s_l\left(t \right)e^{j\frac{2\pi \left( n-1 \right)}{N}l}}}\nonumber\\
                            &+z_m\left(t \right)+x_{\mathrm{J},m}\left(t \right),
\end{align}
where $z_m\left(t \right)\sim \mathcal{C}\mathcal{N} \left( 0,\sigma^2_m \right)$ is the received Gaussian noise at the $m$-th receive element with variance $\sigma^2_m$ and $x_{\mathrm{J},m}\left(t \right)\sim \mathcal{C} \mathcal{N} \left( 0,\sigma^2_{\mathrm{j},m}\right)$ denotes the received jamming signals at the $m$-th element on the receive UCA.

To recover the transmit signal corresponding to the OAM mode $l$, the DFT algorithm is performed on the Eq. (\ref{ym}). Then, the received signal, denoted by $y_{l}\left(t \right)$, can be derived as follows:
\begin{align}\label{yl}
        y_{l}\left(t \right) &= \sum_{m=1}^M y_m\left(t \right) e^{-j\frac{2\pi\left( m-1\right) }{M}l}\nonumber\\
        &= \sqrt{\frac{N}{M}}h_ls_{l}\left(t \right)+\sum_{m=1}^{M}\left( z_m\left(t \right)+x_{\mathrm{J},m}\left(t \right)\right) e^{-j\frac{2\pi\left(m-1 \right) }{M}l}, 
\end{align}
where $h_l$ denotes the channel gain of the OAM mode $l$ from the transmit UCA to the receive UCA and it can be calculated as 
\begin{equation}
h_l=\frac{\beta \lambda \sqrt{N}e^{-j\frac{2\pi \sqrt{d^2+r^2+R^2}}{\lambda}}}{4\pi dj^l}J_l\left( \frac{2\pi rR}{\lambda \sqrt{d^2+r^2+R^2}} \right).
\label{hl}
\end{equation}

In Eq. (\ref{hl}), the $l$-order Bessel function of the first kind, denoted by $J_l\left(\alpha \right) $, is given as follows:
\begin{equation}
J_l\left( \alpha \right) =\frac{j^l}{2\pi}\int_{0}^{2\pi}{e^{jl\tau}e^{j\alpha \sin{\tau}}d\tau}.
\end{equation}
\begin{comment}
内容...We denote by $\tilde{h}_{ml}$ the channel gain from the transmit UCA to the $m$-th element of receive UCA for the OAM mode $l$ which is expressed as
\begin{align}\label{a}
        \tilde{h}_{ml} &=\frac{\beta \lambda \sqrt{N}e^{-j\frac{2\pi \sqrt{d^2+r^2+R^2}}{\lambda}}e^{j\frac{2\pi \left( m-1 \right)}{M}l}}{4\pi dj^l}J_l\left( \frac{2\pi rR}{\lambda \sqrt{d^2+r^2+R^2}} \right)\nonumber\\
        &= h_le^{j\frac{2\pi \left( m-1 \right)}{M}l} ,
\end{align}
where $h_l=\frac{\beta \lambda \sqrt{N}e^{-j\frac{2\pi \sqrt{d^2+r^2+R^2}}{\lambda}}}{4\pi dj^l}J_l\left( \frac{2\pi rR}{\lambda \sqrt{d^2+r^2+R^2}} \right)$ and $J_l\left(\alpha \right) $ denotes the $l$-order Bessel function of the first kind, which is given by
\begin{equation}
J_l\left( \alpha \right) =\frac{j^l}{2\pi}\int_{0}^{2\pi}{e^{jl\tau}e^{j\alpha \sin{\tau}}d\tau}.
\end{equation}
\end{comment}

Since we have already identified the jammed and unjammed OAM modes, we can directly use the OAM modes sets $\mathcal{L}_{\mathrm{j}}$ and $\mathcal{L}_{\mathrm{u}}$ to process the received signals $y_{l}\left(t \right) $. For the jammed OAM modes $\left( l\in \mathcal{L}_{\mathrm{j}} \right) $, energy detection is utilized to recover the transmitted useful information. Next, we compare the average energy $Q$ of the received signal with a predetermined threshold $Q_{\text{th}}$, thereby determining the signal transmitted by the transmitter. 

We perform a $K$-point sampling on this continuous time signal $y_{l}\left(t \right)$ with a sampling interval of $T_s$. From Eq. (\ref{yl}), the signal after sampling is given by
\begin{equation}
y_{l}[k]=\sum_{m=1}^M{y_m[k]e^{-j\frac{2\pi \left( m-1 \right)}{M}l}},
\end{equation} 
where $y_m[k]$ denotes the signal sample from the $m$-th antenna element at the $k$-th sampling point. $y_l\left[ k \right] \sim \mathcal{C} \mathcal{N} \left( 0,\sigma ^2_k \right) $ is a zero-mean complex Gaussian random process. The variance of $y_l[k]$, denoted by $\sigma^2_k$, is calculated by
\begin{equation}
\sigma^2_k=\sigma^2_m+\sigma^2_{\mathrm{j},m}.
\label{var}
\end{equation}

Then, the average energy of the received signal, denoted by $Q$, is given by
\begin{align}
        Q &=\frac{1}{K}\sum_{k=1}^{K}{\left| y_{l}\left[ k \right] \right|^2}= \frac{1}{K}\sum_{k=1}^{K}{\left| \sum_{m=1}^M{y_m[k]e^{-j\frac{2\pi \left( m-1 \right)}{M}l}} \right|^2},
        \label{Q}
\end{align}
where $Q$ is a sum of squares of $K$ independent variables with identical mean and variance. 

By normalizing the variance of $y_l[k]$ according to Eqs. (\ref{var}) and (\ref{Q}), the received average energy satisfies
\begin{equation}
Q/\frac{\sigma^2_k}{2K}=\frac{2K}{\sigma^2_k}Q\sim \chi ^2\left( 2K \right),
\end{equation}
where $\chi ^2\left( 2K \right)$ is the chi-squared distribution with $2K$ degress.

To compare the average energy of the received signals and the energy threshold, we assume that the transmitter sends an on-off keying modulated signal using the length of $I$ preamble sequence. The sequence is defined as $\mathbf{S}=\left[s_1,s_2,\cdots,s_I\right]^{\mathrm{T}}$ where $s_i$ is the $i$-th element and $s_i\in \left\{ 0,1 \right\}$. The indices of elements in the sequence that are 0 and 1 can be represented by the sets $\mathcal{G}_0$ and $\mathcal{G}_1$ respectively, where $\mathcal{G}_0=\left\{ i\left| s_i=0, i=1,2,\cdots ,I \right. \right\} $ and $\mathcal{G}_1=\left\{ i\left| s_i=1, i=1,2,\cdots ,I \right. \right\} $. Therefore, the average power of the 0 and 1 in the preamble sequence is given by
\begin{numcases}{}
	\hat{Q}_0=\frac{1}{I}\sum\limits_{i\in \mathcal{G}_0}{P_i}; \nonumber \\
	\hat{Q}_1=\frac{1}{I}\sum\limits_{i\in \mathcal{G}_1}{P_i},
\end{numcases}
where $P_i$ is the average energy of the $i$-th element in the sequence and it can be expressed as
\begin{equation}
P_i=\frac{1}{K}\sum_{k=1}^K{\left| y_{l}\left[ \left( i-1 \right) K+k \right] \right|^2},1<i<I.
\end{equation}

Consequently, the energy threshold $Q_{\text{th}}$ is given by 
\begin{equation}
Q_{\text{ED}}\left( \hat{Q}_0 ,\hat{Q}_1 \right)
\end{equation}
where $Q_{\text{ED}}$ is the mapping from the $\hat{Q}_b \left(b=0,1\right)$ to $\hat{T}_{th}$.

According to \cite{a10}, this mapping function $Q_{\text{ED}}$ is defined by
\begin{equation}
Q_{\text{ED}}\left( \hat{Q}_0 ,\hat{Q}_1 \right) =\frac{1}{K}\frac{\hat{Q}_0\hat{Q}_1}{\hat{Q}_1-\hat{Q}_0}\ln \left( \frac{p_0}{p_1}\left( \frac{\hat{Q}_1}{\hat{Q}_0} \right) ^K \right),
\end{equation}
where $p_0$ and $p_1$ denote the transmission probabilities of the two symbols, 0 and 1, which can be expressed as $p_0=\frac{1}{I}\left| \mathcal{G}_0 \right|$ and $p_1=\frac{1}{I}\left| \mathcal{G}_1 \right|$, respectively.

Once deriving the average energy $Q$ of the received signal and the energy threshold $\hat{T}_{th}$, we decode the message received by the Rx by comparing the value of $Q$ and $Q_{\text{th}}$. The specific process is given by
\begin{numcases}{}
	Q\ge Q_{\text{th}}, 1; \nonumber \\
	Q<Q_{\text{th}}, 0.
\end{numcases}

When assessing the useful signals transmitted to the receiver on the jammed OAM modes through energy detection, the probability of correct detection, denoted by $P_c$, can be calculated by
\begin{align}\label{Pc}
        P_c =&\left( 1 - F_{\chi^2}\left(Q_{\text{th}}; 2K\right)\right) \mathcal{O}\left( Q\ge Q_{\text{th}}\right)\nonumber\\
                             &+ F_{\chi^2}\left(Q_{\text{th}}; 2K\right)\mathcal{O}\left( Q<Q_{\text{th}}\right),
\end{align}
where $\mathcal{O}\left(\cdot \right) $ is a condition function meaning that when the condition in it is correct, the value of $\mathcal{O}\left(\cdot \right) $ is 1, else is 0. 

In Eq. (\ref{Pc}), the CDF of $Q$, denoted by $F_{\chi^2}\left(Q_{\text{th}}; 2K\right)$, is given as follows:
\begin{equation}
F_{\chi^2}\left(Q_{\text{th}}; 2K\right)=\frac{\delta \left( \frac{2K}{2};\frac{Q_{\text{th}}}{2} \right) }{\Gamma\left(\frac{2K}{2} \right) },
\end{equation}
where $\delta\left( \frac{2K}{2};\frac{Q_{\text{th}}}{2} \right)$ denotes the lower incomplete gamma function and is given by
\begin{equation}
\delta\left( \frac{2K}{2};\frac{Q_{\text{th}}}{2} \right)=\delta\left( K;\frac{Q_{\text{th}}}{2} \right)=\int_{0}^{\frac{Q_{\text{th}}}{2}}x^{K-1}e^{-x}dx.
\end{equation}

To evaluate the performance of the proposed OAM active anti-jamming scheme, we first compute its signal-to-noise (SNR). Based on Eqs. (\ref{Pd}), (\ref{ym}), (\ref{yl}) and (\ref{Pc}), the SNR of the OAM mode $l$ can be calculated by
\begin{comment}
内容...\begin{equation}
\gamma_l=\left\{ \begin{array}{l}
	P_{\text{j}}P_c\frac{\mathbb{E}\left[\frac{\sqrt{N}}{\sqrt{M}}h_{l}s_{l}\left(t \right) \right]}{\mathbb{E}\left[\sum\limits_{m=1}^{M}\left( z_m\left(t \right)+x_{\mathrm{J},m}\left(t \right)\right)e^{-j\frac{2\pi\left(m-1 \right) }{M}l}\right]}, l\in \mathcal{L}_{\mathrm{j}}; \\
	P_{\text{u}}\frac{\mathbb{E}\left[\frac{\sqrt{N}}{\sqrt{M}}h_{l}s_{l}\left(t \right)\right]}{\mathbb{E}\left[\sum\limits_{m=1}^{M}\left( z_m\left(t \right)+x_{\mathrm{J},m}\left(t \right)\right)e^{-j\frac{2\pi\left(m-1 \right) }{M}l}\right]}, l\in \mathcal{L}_{\mathrm{u}}.\\
\end{array} \right. 
\end{equation}
\begin{numcases}{\gamma_l=}
	P_{\text{j}}P_c\frac{\mathbb{E}\left[\frac{\sqrt{N}}{\sqrt{M}}h_{l}s_{l}\left(t \right) \right]}{\mathbb{E}\left[\sum\limits_{m=1}^{M}\left( z_m\left(t \right)+x_{\mathrm{J},m}\left(t \right)\right)e^{-j\frac{2\pi\left(m-1 \right) }{M}l}\right]}, l\in \mathcal{L}_{\mathrm{j}}; \nonumber \\
P_{\text{u}}\frac{\mathbb{E}\left[\frac{\sqrt{N}}{\sqrt{M}}h_{l}s_{l}\left(t \right)\right]}{\mathbb{E}\left[\sum\limits_{m=1}^{M}\left( z_m\left(t \right)+x_{\mathrm{J},m}\left(t \right)\right)e^{-j\frac{2\pi\left(m-1 \right) }{M}l}\right]}, l\in \mathcal{L}_{\mathrm{u}}. 
\label{g}
\end{numcases}
\end{comment}
{\small
\begin{equation}
\gamma_l=\left\{ \begin{array}{l}
	P_{\text{j}}P_c\displaystyle\frac{\mathbb{E}\left[\frac{\sqrt{N}}{\sqrt{M}}h_{l}s_{l}\left(t \right) \right]}{\mathbb{E}\left[\sum\limits_{m=1}^{M}\left( z_m\left(t \right)+x_{\mathrm{J},m}\left(t \right)\right)e^{-j\frac{2\pi\left(m-1 \right) }{M}l}\right]}, l\in \mathcal{L}_{\mathrm{j}}; \\
	P_{\text{u}}\displaystyle\frac{\mathbb{E}\left[\frac{\sqrt{N}}{\sqrt{M}}h_{l}s_{l}\left(t \right)\right]}{\mathbb{E}\left[\sum\limits_{m=1}^{M}\left( z_m\left(t \right)+x_{\mathrm{J},m}\left(t \right)\right)e^{-j\frac{2\pi\left(m-1 \right) }{M}l}\right]}, l\in \mathcal{L}_{\mathrm{u}}.\\
\end{array} \right. 
\end{equation}
}

The SE, denoted by $C$, for our proposed OAM active anti-jamming scheme under hostile jamming can be calculated as follows:
\begin{equation}
C=\sum_{l=\lfloor \frac{2-N}{2} \rfloor}^{ \lfloor \frac{N}{2} \rfloor}{\log _2\left( 1+\gamma_l \right)}.
\end{equation}

\section{Numerical Results}
In this section, numerical results are presented to evaluate the performance of our proposed OAM active anti-jamming scheme. The parameters are set as follows: $r=R=$ 0.75 m, $d=$ 15 m, $M=N$, ${E}_{\text{th}}=0.5$W, $a_0=0.5$, $a_1=2$, the power of jamming signal as 0.1W, and the carrier frequency as 5.8 GHz. 

By comparing the SE with different SNR levels, we analyze the SE of the conventional UCA wireless communication scheme and the proposed OAM active anti-jamming scheme with randomly set jammed OAM modes. In addition, we also analyze the impact of the number of jammed OAM modes on SE for our proposed OAM active anti-jamming scheme.

\begin{figure}[htbp]
%\vspace{-0.5cm}%%调整标题与文字距离间距
%\setlength{\abovecaptionskip}{-0.2cm}%%调整标题与图之间的间距
%\centerline{\includegraphics[scale=0.7]{simulation_t1_6.png}}
\centerline{\includegraphics[scale=0.66]{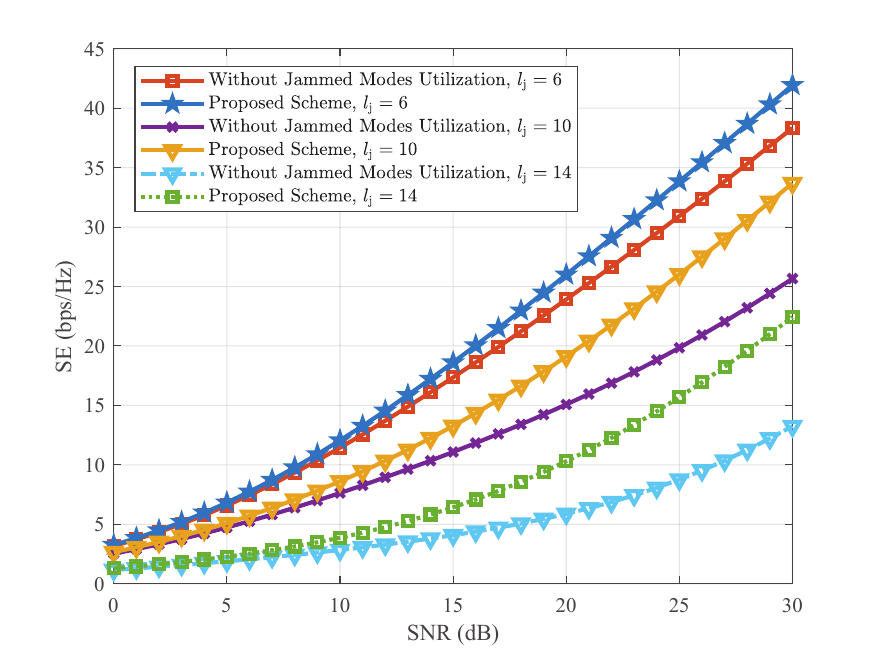}}
\caption{SEs comparison between proposed scheme and jammed UCA system.}
\label{simulation2}
\vspace{-0.2cm}%%调整标题与文字距离间距
\end{figure}

\begin{figure}[htbp]
%\vspace{-0.6cm}%%调整标题与文字距离间距
%\setlength{\abovecaptionskip}{-0.2cm}%%调整标题与图之间的间距
%\centerline{\includegraphics[scale=0.7]{simulation_t2_3.png}}
%\centerline{\includegraphics[scale=0.68]{M_SE.eps}}
\centerline{\includegraphics[scale=0.66]{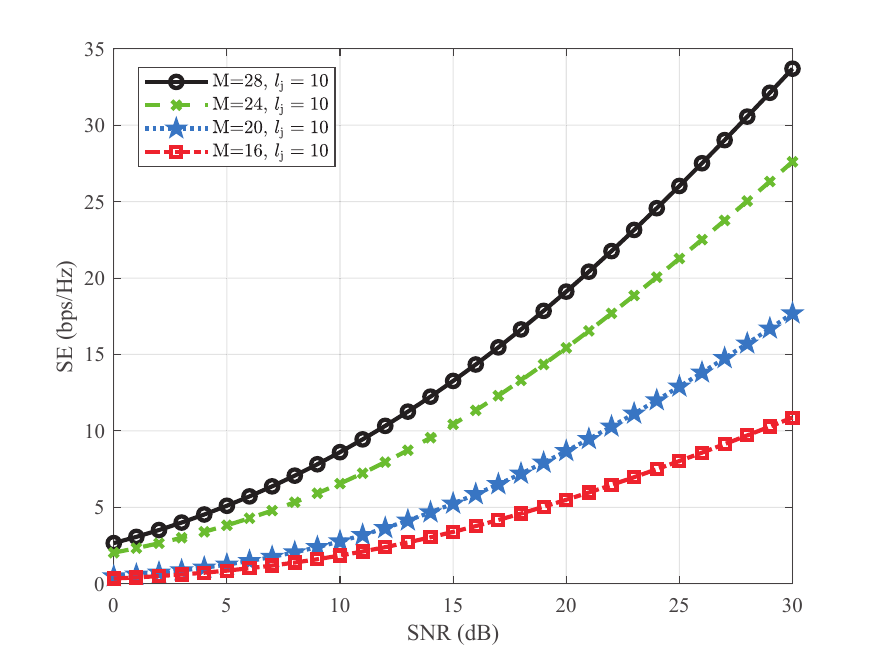}}
\caption{The SEs for proposed OAM active anti-jamming scheme under different $M$.}
\label{simulation}
\vspace{-0.2cm}%%调整标题与文字距离间距
\end{figure}

Figure \ref{simulation2} shows that when the number of jammed OAM modes $l_{\text{j}}$ increases, the SEs of both schemes decrease correspondingly, but the proposed scheme shows higher SE than the conventional scheme under all SNR levels. This is because the proposed scheme increases the SEs of the system by efficiently utilising jammed OAM modes. In conventional schemes, these jammed OAM modes are usually underutilized, leading to a potential loss of SEs. By reallocating the energy resource of jammed and unjammed OAM modes, the proposed scheme not only enhances energy efficiency but also improves the OAM modes utilization efficiency, thus increasing the SEs of the system.

Figure \ref{simulation} depicts the impact of different numbers of transmit and receive elements $\left( M=N=16, 20, 24, 28\right) $ on the SE of wireless communication system when the number of jammed OAM mode $l_{\mathrm{j}}$ is fixed in the proposed OAM active anti-jamming scheme. It is evident that with the increase of SNR, SE has improved in all configurations. This improvement is particularly evident at higher SNR values, as the addition of transmit and receive elements can generate more available OAM modes, thereby increasing spatial multiplexing and enhancing signal processing capabilities in wireless communication systems.

\section{Conclusion}
In this paper, we proposed the OAM active anti-jamming scheme, where the jamming signals were utilized as carriers to send useful information, to improve the anti-jamming performance under co-frequency and co-mode jamming attacks. In particular, OAM-based energy detection was first performed at the transmitter to identify which OAM modes were jammed or unjammed. Then, the transmitter with a PGA was utilized to re-modulate the jammed signals as carriers for transmitting useful information in jammed OAM modes. Meanwhile, with the total transmit power allocated, the unjammed OAM modes were utilized to transmit multiple signals in parallel. Numerical results have shown that our proposed scheme can significantly enhance the SEs as compared with the jammed UCA system.

\bibliographystyle{IEEEtran}
\bibliography{arxiv}

\vspace{12pt}

\end{document}